\documentclass[aps,prd,preprint,showpacs,preprintnumbers,nofootinbib,superscriptaddress]{revtex4}
\usepackage{ae}
\usepackage{bm} 
\usepackage{color}
\usepackage{amssymb}
\usepackage{amsmath}
\usepackage{amsfonts}
\usepackage{graphicx}
\usepackage{slashed}
\usepackage{wasysym}
\usepackage{setspace}
\usepackage{subfigure}
\usepackage{ulem}

\newcommand{\Eqref}[1]{Eq.~\eqref{#1}}

\allowdisplaybreaks

\begin{document}

\setlength{\unitlength}{1mm}
\title{Stimulated photon emission from the vacuum}
\author{Felix Karbstein}\email{felix.karbstein@uni-jena.de}
\affiliation{Helmholtz-Institut Jena, Fr\"obelstieg 3, 07743 Jena, Germany}
\affiliation{Theoretisch-Physikalisches Institut, Abbe Center of Photonics, \\ Friedrich-Schiller-Universit\"at Jena, Max-Wien-Platz 1, 07743 Jena, Germany}
\author{Rashid Shaisultanov}\email{shaisultanov@gmail.com}
\affiliation{Nazarbayev University, National Laboratory Astana, 53 Kabanbay Batyr Avenue, Astana 010000, Republic of Kazakhstan}

\date{\today}

\begin{abstract}
 We study the effect of stimulated photon emission from the vacuum in strong space-time dependent electromagnetic fields.
 We emphasize the viewpoint that the vacuum subjected to macroscopic electromagnetic fields with at least one nonzero electromagnetic field invariant, as, e.g., attainable by superimposing two laser beams, can represent a source term for outgoing photons.
 We believe that this view is particularly intuitive and allows for a straightforward and intuitive study of optical signatures of quantum vacuum nonlinearity in realistic experiments involving the collision of high-intensity laser pulses, and
 exemplify this view for the vacuum subjected to
 a strong standing electromagnetic wave as generated in the focal spot of two counter-propagating, linearly polarized high-intensity laser pulses.
 Focusing on a comparably simple electromagnetic field profile, which should nevertheless capture the essential features of the electromagnetic fields generated in the focal spots of real high-intensity laser beams,
 we provide estimates for emission characteristics and the numbers of emitted photons attainable with present and near future high-intensity laser facilities.
\end{abstract}


\maketitle

\section{Introduction} \label{sec:intro}

The fluctuations of virtual charged particles in the vacuum of quantum electrodynamics (QED) give rise to nonlinear, effective couplings between electromagnetic fields.
While this has been realized theoretically already in the early days of QED \cite{Heisenberg:1935qt,Weisskopf}, the pure electromagnetic nonlinearity of the quantum vacuum still awaits its direct experimental verification on macroscopic scales.  

The advent and planning of high-intensity laser facilities of the petawatt class has triggered a huge interest in ideas and proposals to probe quantum vacuum nonlinearities in realistic all-optical experimental set-ups;
for recent reviews, see \cite{Dittrich:2000zu,Marklund:2008gj,Dunne:2008kc,Heinzl:2008an,DiPiazza:2011tq}.
Typical examples are proposals intended to verify vacuum birefringence \cite{Toll:1952,Baier,BialynickaBirula:1970vy,Adler:1971wn} that can be searched for using macroscopic magnetic fields
\cite{Cantatore:2008zz,Berceau:2011zz} or with the aid of high-intensity lasers \cite{Heinzl:2006xc}, see also \cite{Dinu:2013gaa}.
Alternative concepts suggest the use of time-varying fields and high-precision interferometry \cite{Zavattini:2008cr,Dobrich:2009kd,Grote:2014hja}.
Other commonly studied nonlinear vacuum effects are direct light-by-light scattering \cite{Euler:1935zz,Karplus:1950zz}, photon splitting \cite{Adler:1971wn}, and spontaneous vacuum decay in terms
of Schwinger pair-production in electric fields \cite{Sauter:1931zz,Heisenberg:1935qt,Schwinger:1951nm}.
Further optical signatures of quantum vacuum nonlinearities are those based on interference effects \cite{King:2013am,Tommasini:2010fb,Hatsagortsyan:2011}, photon-photon scattering in the form of laser-pulse collisions \cite{King:2012aw,Lundin:2006wu}, quantum reflection \cite{Gies:2013yxa}, photon merging \cite{Gies:2014jia},
and harmonic generation from laser-driven vacuum \cite{DiPiazza:2005jc,Fedotov:2006ii}. Related effects have also been discussed in the context of searching for minicharged particles \cite{Villalba-Chavez:2013txu}.

In this paper we study the phenomenon of stimulated photon emission from the vacuum in the presence of a strong space-time dependent electromagnetic field (cf. also \cite{Galtsov:1971xm}).
Focusing on a comparably simple electromagnetic field profile, which should nevertheless capture the essential features of the electromagnetic fields generated in the focal spots of real high-intensity laser beams,
we provide estimates for the numbers of emitted photons attainable with present and near future high-intensity laser facilities.

The experimental set-up we have in mind is as follows: A high-intensity laser pulse is split equally into two pulses,  which are separated and directed in a counter-propagation geometry.
Both pulses are focused such that they evolve along the well-defined envelope of a Gaussian beam and their foci overlap. This results in a macroscopic strong-field region about the beam waist (cf. also Fig.~\ref{fig:cartoon}, below).  
The superposition of the two counter-propagating electromagnetic waves results in a standing electromagnetic wave which -- in contrast to a single plane wave -- is characterized by at least one nonzero electromagnetic field invariant.
The idea is to look for induced photons emitted from the strong-field region and to be detected in the field free region. These photons can be considered as emitted from the vacuum subjected to the space-time dependent macroscopic laser field -- whose microscopic composition in terms of laser photons is not resolved --
enabling and stimulating the emission process.
Of course, this scenario can alternatively be interpreted in terms of microscopic laser photon scattering and deflection in the collision of two laser pulses.
From this perspective, the emitted photons correspond to the outgoing photons carrying the imprint of the collision process, i.e., outgoing photons whose properties (in particular their polarization characteristics and propagation directions) differ from the incident laser photons brought into collision.
However, we believe that viewing laser pulse collision processes in terms of a stimulated emission process, i.e., viewing the laser pulses as macroscopic fields, rather than in terms of the constituting laser photons, allows for a particularly intuitive and elegant theoretical treatment.
In this framework it is easy to vary detector sizes and ask for the number of photons carrying the signature of vacuum nonlinearity to be registered in any given solid angle interval, which is not so straightforward in other approaches. In addition, and in contrast to previous studies, e.g., \cite{King:2012aw,Lundin:2006wu}, we can straightforwardly study the polarization properties of the outgoing photons.

Moreover, and from a conceptual level even more important, our approach will also allow us to study photon emission from the vacuum subjected to macroscopic field configurations which are hard to describe as a collection of photons,
like, e.g., rotating inhomogeneous magnetic fields.

Our paper is organized as follows: In Sec.~\ref{sec:calculation} we outline the derivation of the stimulated photon emission rate, and provide explicit analytical results for a particular electromagnetic field configuration,
mimicking the superposition of two counter-propagating laser pulses with the same characteristics. In the diffraction limit these expressions are of a particularly simple form.
Most strikingly, the directional emission characteristics of the induced photons becomes independent of the laser parameters and is described by a generic function.
The number of photons emitted in a specific spherical angle is obtained straightforwardly upon integration of the directional emission characteristics and multiplication with an overall factor determined by the parameters of the used lasers.
Hence, we can easily provide estimates of the number of emitted photons for any desired laser parameters. Section~\ref{seq:Ex+Res} is devoted to the discussion of some explicit results.
We end with conclusions and an outlook in Sec.~\ref{seq:Con+Out}.

\section{Calculation} \label{sec:calculation}

Starting point of our calculation is the one-loop effective Lagrangian in constant external electromagnetic fields (``Heisenberg-Euler effective Lagrangian'') \cite{Heisenberg:1935qt}.
It can be compactly represented as \cite{Schwinger:1951nm} (cf. also \cite{Dittrich:2000zu,Jentschura:2001qr}),
\begin{equation}
 {\cal L}=\frac{e^2}{8\pi^2}\int_{-i\eta}^{\infty-i\eta}\frac{{\rm d}s}{s}\,{\rm e}^{-i(m^2-i\epsilon)s}\left[ab\coth(eas)\cot(ebs)-\frac{a^2-b^2}{3}
 -\frac{1}{(es)^2}\right], \label{eq:effL}
\end{equation}
with $\{\epsilon,\eta\}\to0^+$, elementary charge $e$ and electron mass $m$.
The secular invariants
\begin{equation}
 a=(\sqrt{{\cal F}^2+{\cal G}^2}-{\cal F})^{1/2}\,, \quad b=(\sqrt{{\cal F}^2+{\cal G}^2}+{\cal F})^{1/2}\,,
\end{equation}
are made up of the gauge and Lorentz invariants of the electromagnetic field,
\begin{equation}
 {\cal F}=\frac{1}{4}F_{\mu\nu}F^{\mu\nu}=\frac{1}{2}(\vec{B}^2-\vec{E}^2)\,, \quad {\cal G}=\frac{1}{4}F_{\mu\nu}{}^*F^{\mu\nu}=-\vec{E}\cdot\vec{B}\,. \label{eq:invs}
\end{equation}
Here $^*F^{\mu\nu}=\frac{1}{2}\epsilon^{\mu\nu\alpha\beta}F_{\alpha\beta}$ denotes the dual field strength tensor, and $\epsilon^{\mu\nu\alpha\beta}$ is the totally antisymmetric tensor; $\epsilon^{0123}=1$.
Our metric convention is $g_{\mu \nu}=\mathrm{diag}(-1,+1,+1,+1)$, and we use $c=\hbar=1$.
For completeness note that $ab=\sqrt{{\cal G}^2}$, $a^2-b^2=-2{\cal F}$ and $a^2+b^2=2\sqrt{{\cal F}^2+{\cal G}^2}$.

Strictly speaking, the Heisenberg-Euler Lagrangian~\eqref{eq:effL} describes the effective nonlinear interactions between constant electromagnetic fields mediated by electron-positron fluctuations in the vacuum.
The typical spatial (temporal) extents to be probed by these fluctuations are of the order of the Compton wavelength (time) $\lambda_c=\tau_c=1/m$, with $\lambda_c=3.86\cdot10^{-13}{\rm m}$ and $\tau_c=1.29\cdot10^{-21}{\rm s}$.
Hence, \Eqref{eq:effL} can also be adopted for inhomogeneous electromagnetic fields whose typical spatial (temporal) variation is on scales much larger than the Compton wavelength (time),
i.e., for {\it soft} electromagnetic fields that may locally be approximated by a constant.
Many electromagnetic fields available in the laboratory are compatible with this requirement.
Within the above restrictions, \Eqref{eq:effL} can serve as a starting point to study the effective interaction between dynamical photons and inhomogeneous background electromagnetic fields.

For this purpose it is convenient to decompose the electromagnetic field strength tensor $F^{\mu\nu}$ introduced above as $F^{\mu\nu}\to F^{\mu\nu}(x)+f^{\mu\nu}(x)$ into the field strength tensor of the background field $F^{\mu\nu}(x)$ and the photon field strength tensor $f^{\mu\nu}(x)$ \cite{BialynickaBirula:1970vy}.
To linear order in $f\equiv f^{\mu\nu}$, the Lagrangian can then be compactly written as
\begin{equation}
 {\cal L}=f^{\mu\nu}(x)\frac{\partial{\cal L}}{\partial F^{\mu\nu}}(x)+{\cal O}(f^2)\,. \label{eq:Seff}
\end{equation}
Here we neglected higher-order terms with two or more photons.

Equation~\eqref{eq:effL} is straightforwardly differentiated with respect to $F^{\mu\nu}$, yielding
\begin{equation}
 \frac{\partial{\cal L}}{\partial F^{\mu\nu}}
 =\frac{1}{2}\frac{1}{a^2+b^2}\left[\left(b\frac{\partial{\cal L}}{\partial b}-a\frac{\partial{\cal L}}{\partial a}\right)F_{\mu\nu}
 +{\cal G}\left(\frac{1}{b}\frac{\partial{\cal L}}{\partial b}+\frac{1}{a}\frac{\partial{\cal L}}{\partial a}\right){}^*F_{\mu\nu}\right].
\end{equation}
In particular at leading order in a double expansion of the integrand in \Eqref{eq:effL} in terms of $a$ and $b$
the propertime integral can be performed easily, resulting in
\begin{equation}
 {\cal L}=\frac{e^2}{8\pi^2}\frac{1}{45}\frac{e^2}{m^4}\Bigl[(a^2+b^2)^2+3(ab)^2+ {\cal O}(\varepsilon^6)\Bigr], \label{eq:effLpert}
\end{equation}
with ${\cal O}(a)\sim{\cal O}(b)\sim{\cal O}(\varepsilon)$;
cf. also \cite{Dunne:2004nc} providing the weak field expansion coefficients of the Heisenberg-Euler effective Lagrangian explicitly to all orders.
From \Eqref{eq:effLpert} we obtain the compact expression
\begin{equation}
 \frac{\partial{\cal L}}{\partial F^{\mu\nu}}
 =\frac{e^2}{8\pi^2}\frac{1}{45}\frac{e^2}{m^4}\bigl[4{\cal F}F_{\mu\nu}
 +7{\cal G}{}^*F_{\mu\nu} \bigr] +{\cal O}(\varepsilon^5)\,, \label{eq:delLdelF}
\end{equation}
where we counted $F^{\mu\nu}$ and ${}^*F^{\mu\nu}$ as ${\cal O}(\varepsilon)$.
In our explicit calculations to be performed subsequently for an all-optical laser experiment, we will always limit ourselves to the leading order terms given explicitly in \Eqref{eq:delLdelF}.
As the field strengths attainable in present and near future high-intensity laser facilities are small in comparison to the {\it critical field strength} $E_{\rm cr}\equiv\frac{m^2}{e}$ \cite{Heisenberg:1935qt},
i.e., $\{\frac{eE}{m^2},\frac{eB}{m^2}\}\ll1$, this approximation is well justified.

\begin{figure}
\center
\includegraphics[width=0.3\textwidth]{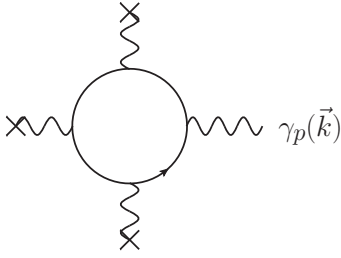}
\caption{Leading order contribution to the stimulated photon emission process in the weak field limit.
The process is cubic in the background field represented by the wiggly lines ending at crosses; cf. Eqs.~\eqref{eq:delLdelF} and \eqref{eq:S}. It results in outgoing photons of wave vector $\vec{k}$ and polarization $p$.}
\label{fig:Feyndiag}
\end{figure}

The amplitude for emission of a single photon with momentum $\vec{k}$ from the vacuum subjected to the background electromagnetic field $F^{\mu\nu}(x)$ is given by
\begin{equation}
 {\cal S}_{(p)}(\vec{k})\equiv\langle\gamma_{p}(\vec{k})|\int{\rm d}^4x\, f^{\mu\nu}(x)\frac{\partial{\cal L}}{\partial F^{\mu\nu}}(x)|0\rangle\,, \label{eq:S}
\end{equation}
with the single photon state denoted by $|\gamma_{p}(\vec{k})\rangle\equiv a^\dag_{\vec{k},p}|0\rangle$ (cf. Fig.~\ref{fig:Feyndiag}). Here $p$ denotes the polarization of the emitted photons.
Representing the photon field in Lorentz gauge as
\begin{equation}
 a^{\mu}(x)=\sum_{p}\int\frac{{\rm d}^3{k}}{(2\pi)^3}\frac{1}{\sqrt{2k^0}}\,\Bigl(\epsilon^{\mu}_{(p)}(k)\,{\rm e}^{-ikx}\,a_{\vec{k},p}+\epsilon^{*\mu}_{(p)}(k)\,{\rm e}^{ikx}\,a^\dag_{\vec{k},p}\Bigl) ,
\end{equation}
where $k^0\equiv|\vec{k}|$, and the sum is over the two physical (transversal) photon polarizations,
we obtain
\begin{equation}
{\cal S}_{(p)}(\vec{k})
 =\frac{i}{\sqrt{2k^0}} \, \hat f^{\mu\nu}_{(p)}(k)\, \frac{\partial{\cal L}}{\partial F^{\mu\nu}}(-k)\,, \label{eq:Sp}
\end{equation}
where $\frac{\partial{\cal L}}{\partial F^{\mu\nu}}(k)=\int{\rm d}^4 x\,{\rm e}^{-ikx}\frac{\partial{\cal L}}{\partial F^{\mu\nu}}(x)$, and we made use of the shorthand notation $\hat f^{\mu\nu}_{(p)}(k)=k^\mu\epsilon^{*\nu}_{(p)}(k)- k^\nu\epsilon^{*\mu}_{(p)}(k)$.

In the vicinity of its beam waist the electromagnetic field of a Gaussian laser beam, corresponding to a fundamental transverse electromagnetic ${\rm TEM}_{00}$ mode, polarized along $\vec{e}_{\rm x}$ and propagating along $\pm\vec{e}_{\rm z}$ can be approximately modeled by the following field configuration
\begin{gather}
 \vec{E}_\pm(x)=\pm{\cal E}\,\vec{e}_{\rm x}\,{\rm e}^{-\frac{4({\rm x}^2+{\rm y}^2)}{w^2}}{\rm e}^{-\frac{{\rm z}^2}{\zeta^2}}\cos\bigl(\Omega(t\mp {\rm z})\bigr), \nonumber\\
 \vec{B}_\pm(x)={\cal E}\,\vec{e}_{\rm y}\,{\rm e}^{-\frac{4({\rm x}^2+{\rm y}^2)}{w^2}}{\rm e}^{-\frac{{\rm z}^2}{\zeta^2}}\cos\bigl(\Omega(t\mp {\rm z})\bigr), \label{eq:beamfields}
\end{gather}
i.e., orthogonal electric and magnetic fields, which -- for given space-time coordinates -- are of the same magnitude, and become maximum for ${\rm x}={\rm y}={\rm z}=0$ (and $t=0$).
Here, we have chosen the orientation of the electric and magnetic fields in such a way that the magnetic field vector points in the same direction ($\vec{e}_{\rm y}$) for both propagation directions $\pm\vec{e}_{\rm z}$;
${\cal E}>0$ denotes the electric/magnetic field amplitude.
The transversal field profile in \Eqref{eq:beamfields} is a Gaussian characterized by its full width $w$ at ${\rm e}^{-1}$ of its maximum.
In longitudinal direction the fields feature a plane-wave type modulation of frequency $\Omega>0$; wavelength $\lambda=\frac{2\pi}{\Omega}$.
Without loss of generality, the beam waist is assumed to be located at $z=0$, such that the Gaussian envelope $\sim{\rm exp}(-{\rm z}^2/\zeta^2)$ can be seen as mimicking the decrease of the field over the Rayleigh range $z_{\rm R}$ which is of the order of $\zeta$.
Note that for real Gaussian beams (for ${\rm x}={\rm y}=0$) the field decrease over the Rayleigh range is described by a Lorentzian profile, which is harder to tackle analytically and thus, would result in less transparent and handy expressions for the vacuum emission probability.
We argue that for our purposes the Gaussian profile captures all relevant features, and -- when providing experimental estimates below -- will actually identify $\zeta=z_{\rm R}$.
Moreover, we neglect diffraction spreading and wavefront curvature effects about the beam waist, arguing that within the Rayleigh range they amount to subleading corrections. Outside the Rayleigh range the fields~\eqref{eq:beamfields} rapidly drop to zero.
High-intensity lasers deliver multicycle pulses of finite duration $\tau$. The pulse duration is also not accounted for explicitly here. Given that $\tau\gg\lambda$, which is typically fulfilled for near infrared high-intensity lasers 
(cf. also Tab.~\ref{tab:lasers}, below) whose pulse duration is $\gtrsim$ tens of femtoseconds and wavelength of the order of $1000$ nanometers, and $\frac{\tau[{\rm fs}]}{\lambda[{\rm nm}]}\approx 300$,
this is justified.
The time scale $\tau$ enters our calculation only as a measure of the interaction time (cf. below).

Let us emphasize that both invariants~\eqref{eq:invs} vanish for a single Gaussian laser beam, modeled by one of the field configurations labeled by $\pm$ in \Eqref{eq:beamfields}.
However, nonzero invariants are attainable by superimposing multiple, e.g., two, Gaussian beams.
Note that macroscopic, non-vanishing invariants could also be realized by a single laser beam if higher laser modes are utilized.
However, in this case the focus area is increased in comparison to the ${\rm TEM}_{00}$ mode and correspondingly the available laser intensity diminished.
Another option is to consider a single Gaussian beam in the limit of a substantial beam divergence $\theta\to\frac{\pi}{2}$ \cite{Monden:2011}, such that wavefront curvature effects become dominant and cannot be neglected; cf. also \cite{Paredes:2014oxa}. Of course, under theses circumstances the laser beam does no longer correspond to a slight modification of a plane-wave like electromagnetic field configuration and both invariants~\eqref{eq:invs} can assume nonzero values, facilitating stimulated photon emission from the vacuum.

\begin{figure}
\center
\includegraphics[width=0.2\textwidth]{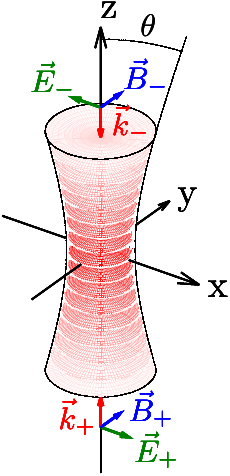}
\caption{Two counter-propagating Gaussian laser beams with wave vectors $\vec{k}_{\pm}=\pm\vec{e}_{\rm z}$ (electric and magnetic field vectors $\vec{E}_\pm$ and $\vec{B}_\pm$)
are superimposed to form a standing electromagnetic wave about the beam focus with nonvanishing field invariant $\cal F$.
The idea is to look for induced photons emitted from this strong-field region, to be detected in the field free region.
For later reference, we also depict the beam divergence $\theta$.}
\label{fig:cartoon}
\end{figure}

At least one invariant can be rendered nonzero by superimposing the two counter-propagating laser beams introduced in \Eqref{eq:beamfields} above.
The resulting electric and magnetic fields amount to standing waves and read
\begin{gather}
 \vec{E}(x)=\vec{E}_+(x)+\vec{E}_-(x)=2{\cal E}\,\vec{e}_{\rm x}\,{\rm e}^{-\frac{4({\rm x}^2+{\rm y}^2)}{w^2}}{\rm e}^{-\frac{{\rm z}^2}{\zeta^2}}\sin(\Omega t)\sin(\Omega {\rm z})\,, \nonumber\\
 \vec{B}(x)=\vec{B}_+(x)+\vec{B}_-(x)=2{\cal E}\,\vec{e}_{\rm y}\,{\rm e}^{-\frac{4({\rm x}^2+{\rm y}^2)}{w^2}}{\rm e}^{-\frac{{\rm z}^2}{\zeta^2}}\cos(\Omega t)\cos(\Omega {\rm z})\,. \label{eq:resfields}
\end{gather}
Figure~\ref{fig:cartoon} is a cartoon of the experimental situation we have in mind.
For the particular electromagnetic field configuration~\eqref{eq:resfields}, the invariants~\eqref{eq:invs} are
\begin{equation}
 {\cal F}(x)={\cal E}^2\,{\rm e}^{-\frac{8({\rm x}^2+{\rm y}^2)}{w^2}}{\rm e}^{-\frac{2{\rm z}^2}{\zeta^2}}\bigl[\cos(2\Omega t)+\cos(2\Omega z)\bigr]\,,\quad {\cal G}=0\,,
\end{equation}
and all components of the field strength tensor $F_{\mu\nu}$ apart from
\begin{equation}
 F_{10}(x)=-F_{01}(x)=\vec{e}_{\rm x}\cdot\vec{E}(x)\,,\quad F_{31}(x)=-F_{13}(x)=\vec{e}_{\rm y}\cdot\vec{B}(x)\,,
\end{equation}
vanish.
Thus, the emission amplitude~\eqref{eq:Sp} can be expressed concisely as 
\begin{equation}
 {\cal S}_{(p)}(\vec{k})
=\frac{2i}{\sqrt{2k^0}}\biggl(\hat f^{10}_{(p)}(k)\,\frac{\partial{\cal L}}{\partial F^{10}}(-k)
 +\hat f^{31}_{(p)}(k)\,\frac{\partial{\cal L}}{\partial F^{31}}(-k)\biggr)\,, \label{eq:Sp1031}
\end{equation}
with
\begin{equation}
 \frac{\partial{\cal L}}{\partial F^{\mu\nu}}(-k)
 =\frac{e^2}{\pi^2}\frac{1}{90}\frac{e^2}{m^4}\int{\rm d}^4x\,{\rm e}^{ikx} {\cal F}(x)F_{\mu\nu}(x) + {\cal O}(\varepsilon^5) \,. \label{eq:delLdelFexpl}
\end{equation}
The Fourier integrals in \Eqref{eq:delLdelFexpl} can be performed straightforwardly.
The integration over time yields $\delta$ functions and the spatial integrations are of Gaussian type.
As the resulting expressions are not very elucidating we do not reproduce them here.

For the following discussion it is convenient to switch to spherical momentum coordinates $\vec{k}={\rm k}\hat{\vec{k}}$, where ${\rm k}=\sqrt{k_{\rm x}^2+k_{\rm y}^2+k_{\rm z}^2}$ and $\hat{\vec{k}}=(\cos\varphi\sin\vartheta,\sin\varphi\sin\vartheta,\cos\vartheta)$,
with $\varphi\in[0\ldots2\pi)$ and $\vartheta\in[0\ldots\pi]$.
The orthogonal vectors to $\vec{k}$ can then be parameterized by a single angle $\beta\in[0\ldots2\pi)$,
\begin{equation}
\vec{e}_{\perp,\beta}=
\left(\begin{array}{c}
  \cos\varphi\cos\vartheta\cos\beta-\sin\varphi\sin\beta \\
  \sin\varphi\cos\vartheta\cos\beta+\cos\varphi\sin\beta \\
  -\sin\vartheta\cos\beta
 \end{array}\right) . \label{eq:e_perpbeta}
\end{equation}
These vectors live in the tangent space of the unit sphere.
Correspondingly, the two transverse polarization modes of photons with wave vector $\vec{k}$ can be spanned by two orthonormalized four-vectors $\epsilon^\mu_{(p)}(\vec{k})$, with $p\in\{1,2\}$,
\begin{equation}
 \epsilon^\mu_{(1)}(\vec{k})=(0,\vec{e}_{\perp,\beta}) \quad\textrm{and}\quad \epsilon^\mu_{(2)}(\vec{k})=\epsilon^\mu_{(1)}(\vec{k})\big|_{\beta\to\beta+\frac{\pi}{2}}\,, \label{eq:epsilons}
\end{equation}
representing linear polarization states in the specific basis characterized by a particular choice of $\beta$. In this work we exclusively focus on linear polarization modes.
Polarizations other than linear can be obtained through linear combinations of the vectors~\eqref{eq:epsilons}.
Resorting to these definitions, the $10$ and $31$ entries of $\hat f^{\mu\nu}_{(p)}(k)$ entering \Eqref{eq:Sp1031} read
\begin{gather}
 \hat f^{10}_{(1)}(k)
  = {\rm k} \bigl(\sin\varphi\sin\beta-\cos\varphi \,u \cos\beta\bigr),
  \quad \hat f^{10}_{(2)}(k)=\hat f^{10}_{(1)}(k)\big|_{\beta\to\beta+\frac{\pi}{2}} , \\
 \hat f^{31}_{(1)}(k)
  = {\rm k}\bigl( \cos\varphi\cos\beta-\sin\varphi \,u \sin\beta \bigr),
  \quad \hat f^{31}_{(2)}(k)=\hat f^{31}_{(1)}(k)\big|_{\beta\to\beta+\frac{\pi}{2}} ,
\end{gather}
where we made use of the shorthand notation $u\equiv\cos\vartheta$.

According to Fermi's golden rule, the number of induced photons with polarization $p\in\{1,2\}$ and momentum ${\rm k}$ in the interval ${\rm dk}$ emitted in the solid angle interval ${\rm d}u {\rm d}\varphi$
is obtained from the modulus squared of \Eqref{eq:Sp} as $\frac{1}{(2\pi)^3}\bigl|{\rm k}\,{\cal S}_{(p)}(\vec{k})\bigr|^2 {\rm dk}\,{\rm d}u\,{\rm d}\varphi$.

A straightforward but somewhat tedious calculation yields the following expressions for the modulus squared of ${\rm k}\,{\cal S}_{(p)}(\vec{k})$,
\begin{multline}
\bigl|{\rm k}\,{\cal S}_{(1)}(u,\varphi)\bigr|^2
 =\frac{\pi^2\,\alpha}{218700}\left(\frac{e{\cal E}}{m^2}\right)^6 \left(\frac{w}{2}m\right)^4\,{\rm k}^3\,\zeta^2\,{\rm e}^{-\frac{\zeta^2\Omega^2+(\frac{w}{2})^2{\rm k}^2}{6}}\,{\rm e}^{-\frac{[\zeta^2-(\frac{w}{2})^2]{\rm k}^2}{6}u^2}  \\
 \times\sum_{n=\pm1}\sum_{l=\pm1}\biggl\{\biggl[
 \sum_{j=0,2} c_j\,{\rm e}^{+\tfrac{lj}{3}\zeta^2{\rm k}\Omega u}(1-u^2)\bigl[\cos(2\beta) + \cos(2\varphi)\bigr] \\
 +\sum_{j=1,3}c_j\,{\rm e}^{+\tfrac{lj}{3}\zeta^2{\rm k}\Omega u}(1-unl)^2[1+\cos(2\varphi-2nl\beta)]  \\
 +{\rm e}^{+\tfrac{l}{3}\zeta^2{\rm k}\Omega u}\,8nl\Bigl[2u[1+\cos(2\varphi)\cos(2\beta)]-(1+u^2) \sin(2\varphi)\sin(2\beta)\Bigr]\biggr]\delta^2({\rm k}-n\Omega)  \\
+\biggl[(1-u^2)\bigl[\cos(2\beta) + \cos(2\varphi)\bigr]\\
+{\rm e}^{+\tfrac{l}{3}\zeta^2{\rm k}\Omega u}(1-unl)^2[1+\cos(2\varphi-2nl\beta)] \biggr]\,\delta^2({\rm k}-3n\Omega) 
     \biggr\}, \label{eq:S^2}
\end{multline}
and $\bigl|{\rm k}\,{\cal S}_{(2)}(u,\varphi)\bigr|^2=\bigl|{\rm k}\,{\cal S}_{(1)}(u,\varphi)\bigr|^2\big|_{\beta\to\beta+\frac{\pi}{2}}$,
with $\alpha=\frac{e^2}{4\pi}$ and coefficients
\begin{equation}
 c_0=4+{\rm e}^{-\frac{4}{3}\zeta^2\Omega^2}, \quad c_1=4+4{\rm e}^{-\frac{2}{3}\zeta^2\Omega^2}\,, \quad c_2= 4{\rm e}^{-\frac{2}{3}\zeta^2\Omega^2}, \quad c_3={\rm e}^{-\frac{4}{3}\zeta^2\Omega^2}. 
\end{equation}
Evidently, only photons with the two distinct frequencies $\omega\in\{\Omega,3\Omega\}$ are induced. 
This is in agreement with elementary physical reasoning: In a Feynman diagrammatic expansion of the effective Lagrangian~\eqref{eq:effL}, the leading terms~\eqref{eq:effLpert} taken into account by us
actually amount to an effective four-photon interaction. Our electromagnetic background field configuration~\eqref{eq:resfields} modeling the counter-propagating laser beams is characterized by a single frequency scale $\Omega$.
Each coupling to the background field configuration can be seen as coupling to a laser photon of frequency $\Omega$. The stimulated emission process involves three laser photons.
Three laser photons can either give rise to a emitted photon of frequency $\Omega$ (two laser photons are scattered into one laser photon and one photon to be emitted) or merge to form a $3\Omega$ photon.  

Hence, upon performing the integration over all possible values of ${\rm k}\in[0\ldots\infty)$ it is convenient to decompose the total number density $\rho_{(p)}(u,\varphi) \equiv \frac{1}{(2\pi)^3}\int_0^\infty{\rm dk}\,|{\rm k}\,S_{(p)}(u,\varphi)|^2$ of induced photons polarized in mode $p$ and emitted in $(u,\varphi)$ direction as
\begin{equation}
 \rho_{(p)}(u,\varphi)=\rho^\Omega_{(p)}(u,\varphi)+\rho^{3\Omega}_{(p)}(u,\varphi),
\end{equation}
where $\rho^{\,\omega}_{(p)}(u,\varphi)$ refers to the number density of induced frequency $\omega$ photons.

These quantities are obtained straightforwardly from \Eqref{eq:S^2}, employing that $\delta^2({\rm k}-\omega)=\frac{\tau}{2\pi}\,\delta({\rm k}-\omega)$, with $\tau$ denoting the time scale of the interaction.
Aiming at the number of photons per laser shot originating from the stimulated emission process, we identify this time scale with the laser pulse duration.

For the $p=1$ polarization mode they read
\begin{equation}
 \rho_{(1)}^{\omega}(u,\varphi)=\left(\frac{e{\cal E}}{m^2}\right)^6 \left(\frac{w}{2}m\right)^4\frac{\Omega \tau}{2\pi}\, h_{(1)}^{\omega}(u,\varphi)\,, \label{eq:rho1}
\end{equation}
where
\begin{multline}
 h_{(1)}^{\Omega}(u,\varphi)=\frac{\alpha}{1749600\pi}(\Omega\zeta)^2\,{\rm e}^{-\frac{[\zeta^2+(\frac{w}{2})^2]\Omega^2}{6}}\,{\rm e}^{-\frac{[\zeta^2-(\frac{w}{2})^2]\Omega^2}{6}u^2}  \\
 \times\sum_{l=\pm1}\biggl\{
 \sum_{j=0,2} c_j\,{\rm e}^{+\tfrac{lj}{3}\zeta^2\Omega^2u}(1-u^2)\bigl[\cos(2\beta) + \cos(2\varphi)\bigr] \\
 +\sum_{j=1,3}c_j\,{\rm e}^{+\tfrac{lj}{3}\zeta^2\Omega^2u}(1-ul)^2[1+\cos(2\varphi-2l\beta)]  \\
 +{\rm e}^{+\tfrac{l}{3}\zeta^2\Omega^2u}\,8l\Bigl[2u[1+\cos(2\varphi)\cos(2\beta)]-(1+u^2) \sin(2\varphi)\sin(2\beta)\Bigr]\biggr\}, \label{eq:rhop1}
\end{multline}
and
\begin{multline}
 h_{(1)}^{3\Omega}(u,\varphi)=\frac{\alpha}{874800\pi}\,3(3\Omega\zeta)^2\,{\rm e}^{-\frac{\zeta^2\Omega^2+(\frac{w}{2})^2(3\Omega)^2}{6}}\,{\rm e}^{-\frac{[\zeta^2-(\frac{w}{2})^2](3\Omega)^2}{6}u^2}  \\
 \times
\biggl\{(1-u^2)\bigl[\cos(2\beta) + \cos(2\varphi)\bigr]\\
+\frac{1}{2}\sum_{l=\pm1}{\rm e}^{+l\zeta^2\Omega^2u}(1-ul)^2[1+\cos(2\varphi-2l\beta)] \biggr\}. \label{eq:rhop2}
\end{multline}
In \Eqref{eq:rho1} we have pulled out an overall factor, such that, apart from $u$ and $\varphi$, the functions $h_{(1)}^{\omega}(u,\varphi)$ only depend on the dimensionless combinations $\xi^2\Omega^2$ and $(\frac{w}{2})^2\Omega^2$.
The results for $p=2$ again follow by shifting the angle $\beta\to\beta+\frac{\pi}{2}$, i.e., $\rho_{(2)}^{\omega}=\rho_{(1)}^{\omega}\big|_{\beta\to\beta+\frac{\pi}{2}}$.

As $\cos(\chi\pm\pi)=-\cos\chi$ and $\sin(\chi\pm\pi)=-\sin\chi$, the {\it total number densities} $\rho^{\,\omega}(u,\varphi)=\sum_{p=1}^2\rho^{\,\omega}_{(p)}(u,\varphi)$
of photons of frequency $\omega$ obtained in a polarization insensitive measurement obviously become independent of $\beta$,
i.e., independent of the specific polarization basis used, as they should:
The resulting expressions are effectively obtained by multiplying Eqs.~\eqref{eq:rhop1} and \eqref{eq:rhop2} with a factor of two and setting all trigonometric functions involving $\beta$ in their arguments to zero.
They read
\begin{equation}
 \rho^{\omega}(u,\varphi)=\left(\frac{e{\cal E}}{m^2}\right)^6 \left(\frac{w}{2}m\right)^4\frac{\Omega \tau}{2\pi}\, h^{\omega}(u,\varphi)\,, \label{eq:rhosum}
\end{equation}
with
\begin{multline}
 h^{\Omega}(u,\varphi)=\frac{\alpha}{874800\pi}(\Omega\zeta)^2\,{\rm e}^{-\frac{[\zeta^2+(\frac{w}{2})^2]\Omega^2}{6}}\,{\rm e}^{-\frac{[\zeta^2-(\frac{w}{2})^2]\Omega^2}{6}u^2}  \\
 \times\sum_{l=\pm1}\biggl\{
 \sum_{j=0,2} c_j\,{\rm e}^{+\tfrac{lj}{3}\zeta^2\Omega^2u}(1-u^2)\cos(2\varphi)
 +\sum_{j=1,3}c_j\,{\rm e}^{+\tfrac{lj}{3}\zeta^2\Omega^2u}(1-ul)^2  \\
 +16\,{\rm e}^{+\tfrac{l}{3}\zeta^2\Omega^2u}\,l u\biggr\}, \label{eq:rhosum1}
\end{multline}
and
\begin{multline}
 h^{3\Omega}(u,\varphi)=\frac{\alpha}{874800\pi}\,3(3\Omega\zeta)^2\,{\rm e}^{-\frac{\zeta^2\Omega^2+(\frac{w}{2})^2(3\Omega)^2}{6}}\,{\rm e}^{-\frac{[\zeta^2-(\frac{w}{2})^2](3\Omega)^2}{6}u^2}  \\
 \times
\biggl\{2(1-u^2)\cos(2\varphi)+\sum_{l=\pm1}{\rm e}^{+l\zeta^2\Omega^2u}(1-ul)^2 \biggr\}. \label{eq:rhosum2}
\end{multline}

The number of photons emitted in a given solid angle interval characterized by $u_2\leq u\leq u_1$ and $\varphi_1\leq\varphi\leq\varphi_2$ is obtained by integration of \Eqref{eq:rho1} or \eqref{eq:rhosum}, respectively.
Note that $\int_{\vartheta_1}^{\vartheta_2}{\rm d}\vartheta\,\sin\vartheta=\int_{u_2}^{u_1}{\rm d}u$, with $u_i=\cos\vartheta_i$ and $i\in\{1,2\}$.

Hence, the total number of frequency $\omega$ photons originating from the stimulated emission process emitted in this solid angle interval ($\Delta u=u_1-u_2$, $\Delta\varphi=\varphi_2-\varphi_1$) is given by
\begin{equation}
 N^{\omega}(\Delta u,\Delta\varphi)=\left(\frac{e{\cal E}}{m^2}\right)^6 \left(\frac{w}{2}m\right)^4\frac{\Omega \tau}{2\pi}\,\int_{u_2}^{u_1}{\rm d}u\int_{\varphi_1}^{\varphi_2}{\rm d}\varphi\, h^{\omega}(u,\varphi)\,, \label{eq:Nsum}
\end{equation}
with $\omega\in\{\Omega,3\Omega\}$.
Obviously, the $\varphi$ integration in \Eqref{eq:Nsum} is trivial. Also the $u$ integration can easily be performed analytically and the result be written in terms of exponential and error functions.
As these results are rather lengthy and do not allow for any additional insights we do not represent them here. 

Analogously, the number of emitted photons polarized in mode $p=1$ is obtained by  
\begin{equation}
 N_{(p)}^{\omega}(\Delta u,\Delta\varphi)=\left(\frac{e{\cal E}}{m^2}\right)^6 \left(\frac{w}{2}m\right)^4\frac{\Omega \tau}{2\pi}\,\int_{u_2}^{u_1}{\rm d}u\int_{\varphi_1}^{\varphi_2}{\rm d}\varphi\, h_{(p)}^{\omega}(u,\varphi)\,. \label{eq:Np}
\end{equation}
As before, the result for the $p=2$ mode follows upon substitution of $\beta\to\beta+\frac{\pi}{2}$.
If the angle parameter $\beta$ is chosen independent of the values of $\varphi$ and $\vartheta$ both integrations can again be performed analytically as for \Eqref{eq:Nsum}. 
However, note that the integrations over the solid angle interval can be significantly complicated if $\beta=\beta(\vartheta,\varphi)$ as is, e.g., necessary if we are interested in all photons polarized perpendicular to $\vec{e}_{\rm x}$; cf. Sec.~\ref{seq:Ex+Res} below.

To maximize the effect of stimulated photon emission, the laser field strength $\cal E$ is preferably rendered as large as possible.
For given laser parameters, $\cal E$ can be maximized by focusing the laser beam down to the diffraction limit, which will be assumed to be the case when providing experimental estimates for the effect below.
The beam diameter of a Gaussian beam of wavelength $\lambda$ focused down to the diffraction limit is given by $w=2\lambda f^\#$ and its Rayleigh range by $z_{\rm R}=\pi\lambda(f^\#)^2$,
with $f^\#$, the so-called $f$-number, defined as the ratio of the focal length and the diameter of the focusing aperture \cite{Siegman}; $f$-numbers as low as $f^\#=1$ can be realized experimentally. 
Recall that in our approximation the length scale $\zeta$ mimics the Rayleigh range $z_{\rm R}$. Correspondingly, aiming at experimental estimates, we identify $\zeta=z_{\rm R}$.

Hence, and perhaps most strikingly, in the diffraction limit the combinations $\zeta^2\Omega^2=(2\pi)^2\pi^2(f^\#)^4$, $(\frac{w}{2})^2\Omega^2=(2\pi)^2(f^\#)^2$
become generic numbers.
In turn, the functions $h_{(p)}^\omega(u,\varphi)$ and $h^\omega(u,\varphi)$ defined in Eqs.~\eqref{eq:rho1}-\eqref{eq:rhosum2} become independent of any explicit laser parameters apart from $f^\#$.
The entire dependence on the laser parameters in Eqs.~\eqref{eq:rho1}, \eqref{eq:rhosum}, \eqref{eq:Nsum} and \eqref{eq:Np} is encoded in the overall prefactor 
\begin{equation}
 \left(\frac{e{\cal E}}{m^2}\right)^6 \left(\frac{w}{2}m\right)^4\frac{\Omega\tau}{2\pi} \ \ \xrightarrow{\text{diffraction\ limit}} \ \
 (f^\#)^4\left(\frac{e{\cal E}}{m^2}\right)^6 \left(\frac{\lambda}{\lambda_c}\right)^3\frac{\tau}{\tau_c}=(f^\#)^4\,2^3\left(\frac{I}{I_{\rm cr}}\frac{\lambda}{\lambda_c}\right)^3\frac{\tau}{\tau_c}\,,
\end{equation}
where $I=\frac{1}{2}{\cal E}^2$ denotes the mean intensity per laser beam and $I_{\rm cr}\equiv(\frac{m^2}{e})^2=4.68\cdot10^{29}\frac{\rm W}{{\rm cm}^2}$ is the {\it critical intensity}.
Moreover, $\lambda_c$ and $\tau_c$ are the Compton wavelength and time introduced above.

\section{Results and Discussion} \label{seq:Ex+Res}

Here, we aim at providing some rough estimates of the number of photons resulting from the stimulated photon emission process.
To this end we assume the original multicycle laser pulse characterized by its wavelength $\lambda$, pulse energy $W$ and pulse duration $\tau$
to be split into two counter-propagating pulses of energy $W/2$ to be focused down to the diffraction limit with $f^\#=1$,
and give the numbers of emitted photons per shot. The experimental scenario is sketched in Fig.~\ref{fig:cartoon}.

The counter-propagating laser pulses are superimposed to form a standing electromagnetic wave within their overlapping foci; cf. \Eqref{eq:resfields} above.
Assuming Gaussian beams, the effective focus area is conventionally defined to contain $86\%$ of the beam energy ($1/e^2$ criterion for the intensity).
Correspondingly, the mean intensity for each beam is estimated as
\begin{equation}
 I\approx \frac{0.86\,(W/2)}{\tau\,\sigma}\,,
\label{eq:Ipump}
\end{equation}
with focus area $\sigma\approx\pi\lambda^2$.
For completeness, also note that the divergence $\theta$ of a Gaussian beam in the considered limit is given by $\theta=\frac{1}{\pi}$ \cite{Siegman} (cf. Fig.~\ref{fig:cartoon}).
Therewith, all physical parameters in Eqs.~\eqref{eq:Nsum} and \eqref{eq:Np} are specified and the number of emitted photons can be evaluated. 
As $W$ is conventionally given in units of joules, $\tau$ in femtoseconds and $\lambda$ in nanometers, it is helpful to note that
\begin{equation}
  \left(\frac{I}{I_{\rm cr}}\frac{\lambda}{\lambda_c}\right)^3\frac{\tau}{\tau_c}\approx 3.40\cdot10^{11}\left(\frac{W[J]}{\tau[{\rm fs}]\,\lambda[{\rm nm}]}\right)^3 \tau[{\rm fs}]\,. \label{eq:J-fs-nm}
\end{equation}

\begin{figure}[h]
\center
\subfigure{\includegraphics[width=0.4\textwidth]{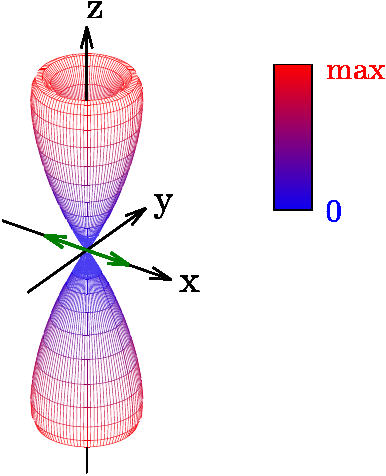}}\hspace*{0.2cm}
\subfigure{\includegraphics[width=0.485\textwidth]{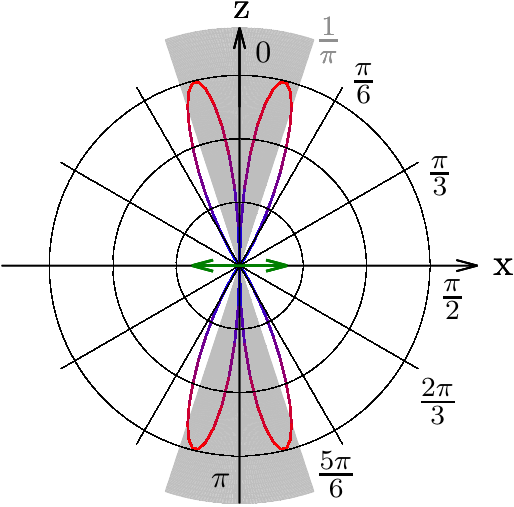}}
\caption{{\bf Left:} Directional emission characteristics $h^\omega(\cos\vartheta,\varphi)\sin\vartheta$ for a polarization insensitive measurement of the photons emitted from the laser focus in arbitrary units.
The result exhibits a superficial rotational symmetry about the beam axis $\rm z$ (cf. Fig.~\ref{fig:cartoon}) and a mirror symmetry with respect to the $\rm x$-$\rm y$ plane.
Deviations from the rotational symmetry $\sim\cos(2\varphi)$ [cf. \Eqref{eq:rhosum1}] are extremely tiny and indiscernible here.
The electric field and thus, the polarization vector of the electromagnetic field configuration~\eqref{eq:resfields} in the laser focus oscillates along $\vec{e}_{\rm x}$ (green arrows).
{\bf Right:} Cut through the three dimensional emission characteristics (left) in the $\rm x$-$\rm z$ plane.
For comparison, the divergence $\theta=\frac{1}{\pi}$ of a Gaussian beam focused down to the diffraction limit is depicted in gray, i.e., a Gaussian beam encompasses photons propagating in all gray shaded directions.
While most photons are emitted into these directions, a certain fraction is emitted into directions outside the laser beam.
Practically no photons are emitted under angles $>\frac{\pi}{6}=30^\circ$ about the beam axis.}
\label{fig:Ntotal}
\end{figure}

Before providing some explicit estimates of the numbers of photons resulting from the stimulated emission process attainable with present and near future high-intensity laser facilities,
we focus on the directional emission characteristics encoded in the functions $h_{(p)}^\omega(u,\varphi)$.
Let us emphasize again that -- in the diffraction limit, and particularly for $f^\#=1$ -- these characteristics are independent of the actual laser parameters, and thus, are the same for all lasers.
For a polarization insensitive measurement of the emitted photons the relevant directional emission characteristics as a function $\varphi,\vartheta$ are described by $h^\omega(\cos\vartheta,\varphi)\sin\vartheta$; recall that $|\frac{{\rm d}u}{{\rm d}\vartheta}|=\sin\vartheta$.
We depict them in Fig.~\ref{fig:Ntotal}.
The total number $N^\omega_{\rm tot}$ of emitted photons of frequency $\omega\in\{\Omega,3\Omega\}$ is obtained straightforwardly from \Eqref{eq:Nsum} with $u_2=-1$, $u_1=1$, $\varphi_1=0$ and $\varphi_2=2\pi$.
This results in
\begin{equation}
 \left\{\begin{array}{c}
  \! N^\Omega_{\rm tot}\\
  \! N^{3\Omega}_{\rm tot}
 \end{array}\right\}
 \approx
 \left\{\begin{array}{c}
   2.94\cdot10^{7} \\
   8.48\cdot10^{-16}
 \end{array}\right\}\left(\frac{W[J]}{\lambda[{\rm nm}]}\right)^3\!\left(\frac{1}{\tau[{\rm fs}]}\right)^2 . \label{eq:res1}
\end{equation}
As the $3\Omega$ signal is severely suppressed, we do not study it any further in the remainder of this paper.

It is instructive to also provide the total number of photons of frequency $\Omega$ emitted into directions outside the laser beam, to be denoted by $N^\Omega_{{\rm tot},>\theta}$ (cf. Fig.~\ref{fig:Ntotal})
and given by  
\begin{equation}
 N^\Omega_{{\rm tot},>\theta}\approx 9.84\cdot10^{6}\left(\frac{W[J]}{\lambda[{\rm nm}]}\right)^3\!\left(\frac{1}{\tau[{\rm fs}]}\right)^2. \label{eq:res2}
\end{equation}

As the laser field is polarized along $\vec{e}_{\rm x}$, it is particularly interesting to ask for the number of emitted photons with perpendicular polarization, fulfilling 
$\vec{e}_{\perp,\beta}\cdot\vec{e}_{\rm x}=0$ $\leftrightarrow$ $\beta=\arctan(\cot\varphi\cos\vartheta)$; cf. Eqs.~\eqref{eq:e_perpbeta} and \eqref{eq:epsilons} above.
Hence, to project out the emitted photons with polarization vector in the $\rm y$-$\rm z$ plane, the angle parameter $\beta$ has to be adjusted as a function of the emission direction parameterized by the angles $\varphi$ and $\vartheta$.
With regard to Eqs.~\eqref{eq:rhop1} and \eqref{eq:rhop2} it is helpful to note that $\sin(2\arctan\chi)=\frac{2\chi}{1+\chi^2}$,
while $\cos(2\varphi-2l\arctan\chi)=\frac{1-\chi^2}{1+\chi^2}\cos(2\varphi)+2l\frac{\chi}{1+\chi^2}\sin(2\varphi)$ for $l=\pm1$.

\begin{figure}[h]
\center
\subfigure{\includegraphics[width=0.465\textwidth]{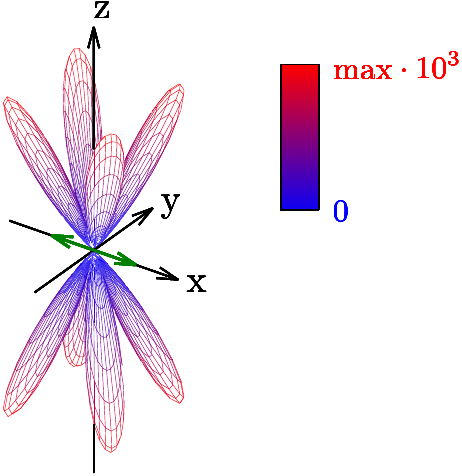}}\hspace*{-1.2cm}
\subfigure{\includegraphics[width=0.515\textwidth]{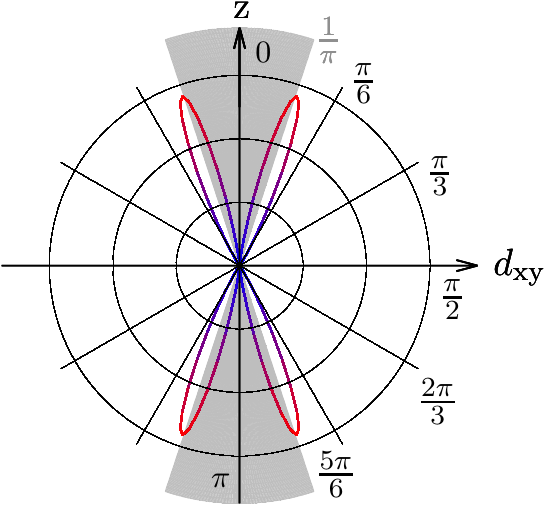}}
\caption{{\bf Left:} Directional emission characteristics $h_{\perp}^\omega(\cos\vartheta,\varphi)\sin\vartheta$ for the emitted photons polarized perpendicular to the laser field (polarized in $\rm x$ direction) in arbitrary units; cf. also Fig.~\ref{fig:Ntotal}.
For completeness, we note that the maximum value in this plot amounts to one thousandth of the maximum value in Fig.~\ref{fig:Ntotal},
i.e., the emission signal for perpendicular polarized photons is significantly smaller than the result obtained when including all polarizations.
The three dimensional emission characteristics is symmetric with respect to the coordinate planes.
{\bf Right:} Cut through the three dimensional figure (left) along the $d_{\rm xy}$-$\rm z$ plane, where $d_{\rm xy}$ is the diagonal in the $\rm x$-$\rm y$ plane above which the signal becomes maximum.
}
\label{fig:Nperpx}
\end{figure}

Defining $h_{\perp}^\omega(u,\varphi)\equiv h_{(1)}^\omega(u,\varphi)\big|_{\beta=\arctan(\cot\varphi\cos\vartheta)}$,
the directional emission characteristics for photons with polarization vector perpendicular to the polarization direction of the laser $\vec{e}_{\rm x}$ are described by
$h_{\perp}^\omega(\cos\vartheta,\varphi)\sin\vartheta$.
We depict them in Fig.~\ref{fig:Nperpx}.
The number of frequency $\Omega$ photons polarized perpendicular to $\vec{e}_x$ and emitted in the solid angle interval parameterized by $u_1\leq u\leq u_2$ and $\varphi_1\leq\varphi\leq\varphi_2$ is obtained by integration of $h_{\perp}^\omega(u,\varphi)$ (cf. Sec.~\ref{sec:calculation} above).
Integrating over the full solid angle results in $N^\Omega_{\perp}$, and just integrating over all directions outside the laser beam in $N^\Omega_{\perp,>\theta}$.
Our explicit results are
\begin{equation}
 \left\{\begin{array}{c}
  \! N^\Omega_{\perp}\\
  \! N^{\Omega}_{\perp,>\theta}
 \end{array}\right\}
 \approx
 \left\{\begin{array}{c}
   1.35\cdot10^{4} \\
   1.18\cdot10^{4}
 \end{array}\right\}\left(\frac{W[J]}{\lambda[{\rm nm}]}\right)^3\!\left(\frac{1}{\tau[{\rm fs}]}\right)^2 . \label{eq:Nperp}
\end{equation}
Note that these numbers are about a factor of $10^{-3}$ smaller than those for all polarizations given in the first line of \Eqref{eq:res1} and \Eqref{eq:res2}; cf. also Figs.~\ref{fig:Ntotal} and \ref{fig:Nperpx}.
In Tab.~\ref{tab:lasers} we list some explicit estimates for the numbers~\eqref{eq:res1}-\eqref{eq:Nperp} of photons of frequency $\Omega$ originating from the stimulated emission process for various present and near future high-intensity laser facilities.
\begin{table}[h]
    \begin{tabular}{ | c || c | c | c || c | c | c | c |}
    \hline
    Laser & \ $W$[J]\ \ & \ $\tau$[fs]\ \ & \ $\lambda$[nm]\ \ & $N^\Omega_{\rm tot}$ & $N^\Omega_{{\rm tot},>\theta}$ & $N^\Omega_\perp$ & $N^\Omega_{\perp,>\theta}$ \\
    \hline
    \hline
    POLARIS & $150$ & $150$ & $1030$ & $4.04$ & $1.35$ & \ $1.86\cdot10^{-3}$\ \ & \ $1.62\cdot10^{-3}$ \ \ \\
    \hline
    Vulcan & $500$ & $500$ & $1054$ & $12.6$ & $4.22$ & $5.77\cdot10^{-3}$ & $5.04\cdot10^{-3}$ \\
    \hline
    Omega EP & $2000$ & $10000$ & $1054$ & $2.01$ & $6.75\cdot10^{-1}$ & $9.20\cdot10^{-4}$ & $8.08\cdot10^{-4}$ \\
    \hline
    ELI Prague & $1500$ & $150$ & $1054$ & \ $3.77\cdot10^3$\ \ & \ $1.26\cdot10^3$\ \ & $1.73$ & $1.51$ \\
    \hline
    ELI-NP & $2\,\times\,250$ & $25$ & $800$ & $1.15\cdot10^4$ & $3.86\cdot10^3$ & $5.28$ & $4.62$ \\
    \hline
    XCELS & \ $12\,\times\,400$\ \ & $25$ & $910$ & $6.90\cdot10^6$ & $2.32\cdot10^6$ & $3.17\cdot10^3$ & $2.78\cdot10^3$ \\
    \hline
    \end{tabular}
 \caption{Numbers of induced photons resulting from the stimulated vacuum emission process for various present and near future high-intensity laser systems, characterized by their pulse energy $W$, pulse duration $\tau$ and wavelength $\lambda$.
 Apart from the total numbers of frequency $\Omega$ photons emitted in all directions, $N^\Omega_{\rm tot}$, and in all directions outside the laser beam, $N^\Omega_{{\rm tot},>\theta}$,
 we provide the numbers of emitted photons polarized perpendicular to the initial laser field, $N^\Omega_\perp$ and $N^\Omega_{\perp,>\theta}$.}
 \label{tab:lasers}
\end{table}

However, let us emphasize that only those photons emitted in the $\rm y$-$\rm z$ plane ($\vartheta=\frac{\pi}{2}$) can be polarized in the same direction as the original laser beam.
Only here, the polarization vectors which live in the tangent space of the unit sphere [cf. \Eqref{eq:e_perpbeta}] can point in the $\vec{e}_{\rm x}$ direction.
With regard to the total number of emitted photons, these photons amount to a negligible fraction:
This becomes particularly obvious when looking at the directional emission characteristics for the total number of photons depicted in Fig.~\ref{fig:Ntotal} (left).
All photons that might have their polarization vector in the same direction as the original laser field lie on the intersection of the $\rm y$-$\rm z$ plane with the three dimensional emission characteristics.
Clearly, their contribution to the integral~\eqref{eq:Nsum} yielding the number of emitted photon number in three dimensions is negligible as it is to be multiplied with ${\rm d}\varphi\to0$ when performing the integration over any solid angle interval.  
In all other emission directions ($\vartheta\neq\frac{\pi}{2}$) the induced photons originating from the stimulated emission process are polarized differently than the laser,
implying that basically all emitted photons are polarized differently than the laser beam triggering the effect.

In turn, this could be used to distinguish the signal (emitted photons) from the laser photons of the same frequency.
For example, equipping a photon detector with a polarizer blocking the polarization of the laser beam along $\vec{e}_{\rm x}$ still a significant fraction of the total numbers of photons emitted in directions outside the laser beam, $N^\Omega_{{\rm tot},>}$, should be detectable: All photons with a nonvanishing polarization component perpendicular to $\vec{e}_{\rm x}$ will actually contribute to the signal.
With respect to such measurement, our result~\eqref{eq:Nperp} for the truly perpendicular polarized emission signal $N^\Omega_{\perp,>}$ (cf. also Fig.~\ref{fig:Nperpx} and Tab.~\ref{tab:lasers}) just amounts to the absolute minimum number of emitted photons to be detected.

\section{Conclusions and Outlook} \label{seq:Con+Out}

In this paper we have studied and interpreted a specific laser pulse collision process in terms of  stimulated single photon emission from the vacuum in strong space-time dependent electromagnetic fields.
More specifically, we have focused on a particular field configuration mimicking the electromagnetic field in the focal spot of two counter-propagating, linearly polarized high-intensity laser beams with their polarization vectors pointing in the same direction.

It would be interesting to extend our study to other electromagnetic field configurations attainable in the overlapping foci of two high-intensity laser pulses,
e.g., to deviate from the counter-propagation geometry by letting the beams collide under an relative angle and to study other laser polarizations.
Moreover, the electromagnetic field profiles to mimic the laser beams should eventually be improved to account for more features of real, experimentally attainable pulses.
In particular the Gaussian profile mimicking the finite Rayleigh length in the present study should be replaced by a Lorentzian profile.
Besides, in a latter step of this program also a dedicated detection set-up should be worked out and the precise numbers of the detectable photons originating from the stimulated emission process should be specified. 
Let us emphasize again that in the present study we rather intended to underpin our viewpoint of interpreting the vacuum subjected to macroscopic strong electromagnetic (laser) fields as source term for outgoing photons. To this end, we present first estimates of the photon numbers attainable from the effect of stimulated photon emission in an all-optical experimental set-up within this framework.

Finally, and perhaps most importantly, the approach adopted by us can be straightforwardly extended to processes involving $n\geq2$ external photons, with $n\in\mathbb{N}$,
attainable by expanding the effective Lagrangian~\eqref{eq:effL} with $F^{\mu\nu}(x)\to F^{\mu\nu}(x)+f^{\mu\nu}(x)$ to ${\cal O}(f^n)$;
cf. our discussion in the context of \Eqref{eq:Seff} above, and also \cite{Karbstein:2015cpa}.

\section*{Acknowledgments}

The authors thank H.~Gies for many stimulating discussions.
FK is grateful to Matt~Zepf for various helpful and enlightening discussions.
FK acknowledges support by the DFG (SFB-TR18).
RS acknowledges support by the Ministry of Education and Science of the Republic of Kazakhstan.

\end{document}